\documentclass[%
 reprint,
nofootinbib,
nobibnotes,
 amsmath,amssymb,
 aps, prd,
 longbibliography,
floatfix,
]{revtex4-2}

\usepackage{graphicx}
\usepackage{dcolumn}
\usepackage{bm}
\usepackage[normalem]{ulem}
\usepackage[english]{babel}
\usepackage{fancyhdr}
\usepackage{amsmath}
\usepackage{amssymb}
\usepackage{amsfonts}
\usepackage{psfrag}
\usepackage[utf8]{inputenc}
\usepackage[bf,footnotesize,justification=Justified,format=plain]{caption}
\usepackage[dvipsnames]{xcolor}
\usepackage[colorlinks=True, citecolor=blue, linkcolor=blue, urlcolor=blue,linktocpage]{hyperref}
\usepackage{amsthm}
\usepackage{gensymb}
\usepackage{subfig}
\usepackage{physics}
\usepackage{array}
\usepackage{tcolorbox, mathtools}
\usepackage{soul}
\tcbuselibrary{skins}
\newtcolorbox{mybox}[1][]{before=\centering, drop fuzzy shadow, enhanced, colframe=blue, fonttitle=\bfseries, title=#1, center title}
\usepackage{aas_macros}
\usepackage{cleveref}

\newcommand{\sch}{Schwarzschild }

\newcommand{\centra}{CENTRA, Departamento de Física, Instituto Superior Técnico – IST,
Universidade de Lisboa – UL, Avenida Rovisco Pais 1, 1049-001 Lisboa, Portugal}

\begin{document}

\title{Vanishing of all redshift modes in \sch ringdown}

\author{Adrien Kuntz}
 \email{adrien.kuntz@tecnico.ulisboa.pt}
\affiliation{\centra}
\author{Matteo Della Rocca}
\email{matteo.dellarocca@uniroma1.it}
\affiliation{Dipartimento di Fisica, Sapienza Universit\`a di Roma, Piazzale Aldo Moro 5, 00185, Roma, Italy}
\affiliation{INFN, Sezione di Roma, Piazzale Aldo Moro 2, 00185, Roma, Italy}

\begin{abstract}
Several studies of black hole ringdown from particles plunging into black holes have identified contributions decaying at integer multiples of the surface gravity, called redshift modes, horizon modes, and direct waves. We show that, for Schwarzschild black holes, every one of these contributions has vanishing amplitude in the observable waveform. The cancellation follows from causality, which forces the source-integrated Green function to vanish on the light cone. Individual quasi-normal mode overtones still carry non-zero redshift-mode contributions, but these cancel exactly once the sum over overtones is performed; the so-called impulsive contribution to the waveform acts precisely as the counterterm enforcing this cancellation. Finally, we provide a motivation to the standard regularization of quasinormal mode excitation coefficients since divergences give rise to vanishing redshift modes. 
\end{abstract}

\date{\today}

\maketitle

\section{Introduction}

The direct detection of gravitational waves (GWs) from coalescing black holes (BHs) by the LIGO–Virgo–KAGRA (LVK) collaboration~\cite{Abbott:2020niy,KAGRA:2021vkt,LIGOScientific:2025slb} has opened an unprecedented observational window onto the strong-field regime of gravity. Among the various stages of a binary BH coalescence, the final relaxation of the remnant object — the so-called \emph{ringdown} phase — plays a particularly important role. In this regime, the newly formed BH settles down toward equilibrium through a superposition of damped oscillations known as quasi-normal modes (QNMs), whose frequencies and damping times are entirely determined, within General Relativity, by the mass and spin of the remnant BH~\cite{Berti:2025hly}.

Significant progress has recently been achieved in understanding the physical content of waveforms in the merger-ringdown regime. Although phenomenological ringdown models typically describe QNM amplitudes as constant in time, linear perturbation theory predicts instead that these amplitudes build up dynamically during the merger before asymptoting to constant values at late times~\cite{Lagos_2023,Chavda:2024awq}. Building on recent developments concerning the Green function of \sch BHs~\cite{Kuntz:2025gdq,Arnaudo:2025uos,Arnaudo:2025kit,Su:2026fvj,Aruquipa:2026tga,Arnaudo:2026tcy}, these time-dependent QNM amplitudes were computed for the first time for a plunging point particle in \sch spacetime in Refs.~\cite{DeAmicis:2025xuh,DeAmicis:2026wqd}.

An intriguing feature emerged from these computations. While the amplitude of the fundamental mode converges to a constant at late times, the amplitudes associated with overtones exhibit a secular divergence. As shown in Ref.~\cite{DeAmicis:2025xuh}, when combined with the standard exponential decay of the QNM contribution, this divergence gives rise to additional components in the waveform, dubbed \emph{redshift modes}. These modes decay exponentially with a rate proportional to the BH surface gravity $\kappa_h = 1/4M$, where $M$ denotes the BH mass, but unlike ordinary QNMs they do not oscillate in time. Should such components be physically present in gravitational-wave signals, they would provide an especially interesting probe of near-horizon physics, since they are directly associated with the geometry of the horizon rather than with the light-ring structure governing ordinary QNMs.

Independently of these developments, several other features of BH waveforms and Green functions closely related to redshift modes have been investigated in the literature. The first concerns the so-called \emph{horizon modes} appearing in the waveform of rotating BHs~\cite{Mino:2008at,Zimmerman:2011dx}. These modes originate from an additional pole in the Laplace transform of the waveform, located at a frequency determined by the asymptotic motion of the plunging particle near the horizon. In the \sch case, this frequency coincides precisely with the redshift mode frequency.
A second related structure is provided by the \emph{Matsubara modes}, namely the frequencies at which the ingoing wavefunction at the horizon (and thus its ingoing amplitude at infinity, $A_{\rm in}$) develop poles~\cite{Arnaudo:2025uos,Motohashi:2026mbn}, despite the fact that the full Green function itself remains regular there~\cite{Kuntz:2025gdq}. These poles occur exactly at the redshift mode frequencies~\cite{Arnaudo:2025uos,Kuntz:2025gdq,Motohashi:2026mbn}. However, at the moment there is no claim that these modes can be observed in waveforms. 
Finally, Ref.~\cite{Oshita:2025qmn} identified a further contribution to the waveform, referred to as the \emph{direct wave}, obtained through a stationary-phase analysis of the radiation emitted by a plunging particle. Once again, the corresponding frequency asymptotically approaches the redshift-mode frequency, although it was argued in Ref.~\cite{Oshita:2025qmn} that this contribution becomes screened because of the pole structure of the ingoing amplitude $A_{\rm in}$ at the Matsubara frequencies.
These various contributions to the waveform have been derived using rather different methods and approximations, which obscures their mutual relation. Since they all share the same decay rate, an immediate question is whether they in fact originate from the same physical mechanism. More concretely, what is their respective contribution to the observable waveform? In this work, we demonstrate that \emph{all} of these components — redshift modes, horizon modes, and direct waves — possess a vanishing amplitude in the waveform generated by particles plunging into a \sch BH, and are therefore unobservable. We show that this remarkable cancellation is ultimately a consequence of causality, which implies that the Green function integrated against the source vanishes when evaluated on the light cone.

Our conclusions are fully consistent with the results of Refs.~\cite{DeAmicis:2025xuh,DeAmicis:2026wqd}. Indeed, each individual overtone still contains a nonvanishing redshift-mode contribution; however, these contributions cancel exactly once the full sum over overtones is performed. Although we do not yet possess a rigorous proof extending this cancellation to rotating BHs, we expect the result to persist in the Kerr case, since the essential ingredient underlying the proof is causality itself.

We further investigate the relation between redshift modes and QNM excitation coefficients, namely the asymptotic late-time amplitudes of QNMs. These coefficients were computed in earlier works~\cite{Leaver:1986gd,Hadar:2009ip,Berti_2006,Berti:2006wq,Zhang:2013ksa,Sun:1988tz,DellaRocca:2025zbe}, although the corresponding amplitudes are formally divergent and therefore require a regularization prescription. We explicitly show that the  procedures used in Refs.~\cite{Leaver:1986gd,Hadar:2009ip,Berti_2006,Berti:2006wq,Zhang:2013ksa,Sun:1988tz,DellaRocca:2025zbe} corresponds to discard the redshift modes from the waveform. Our proof of the vanishing of redshift modes therefore provides a firm theoretical justification for the regularization procedures adopted in Refs.~\cite{Leaver:1986gd,Hadar:2009ip,Berti_2006,Berti:2006wq,Zhang:2013ksa,Sun:1988tz,DellaRocca:2025zbe}. More generally, our results suggest a natural definition of regularized time-dependent QNM amplitudes, obtained by subtracting from each overtone amplitude its associated secularly divergent redshift-mode contribution.

The remainder of this paper is organized as follows. In Section~\ref{sec:setup}, we introduce the formalism underlying our analysis and derive the expression for the waveform generated by a point particle plunging into a \sch BH. In Section~\ref{sec:LO}, we show that the leading-order redshift mode, decaying as $e^{-u/2}$, has vanishing amplitude. In Section~\ref{sec:proofAllOrders}, we extend this result to all higher-order redshift modes. Section~\ref{sec:excitationCoeffs} is devoted to the relation between redshift modes and the regularization of excitation coefficients. Finally, in Section~\ref{sec:CL}, we summarize our results and discuss their implications. Throughout this paper we use units $G=c=1$, and we set the mass of the \sch BH to $M=1/2$, so that the horizon is located at $r=1$ in \sch coordinates.

\section{Setup} \label{sec:setup}

\subsection{Infall into a \sch BH} \label{sec:infall}

Geodesics of \sch spacetime are characterized by an energy $E$ and angular momentum $L$, with a velocity 
\begin{align} \label{eq:veloc}
    v_\mathrm{P} = \frac{\mathrm{d} x_\mathrm{P}}{\mathrm{d} t} &= - \frac{1}{E} \sqrt{E^2 - A(r_\mathrm{P}) \bigg( \frac{L^2}{r_\mathrm{P}^2}+1 \bigg)} \; , \\
    \frac{\mathrm{d} \phi_\mathrm{P}}{\mathrm{d} t} &= \frac{L A(r_\mathrm{P})}{E r_\mathrm{P}^2} \; ,
\end{align}
where $r_\mathrm{P}$ is the point-particle location, $A(r) = 1-1/r$ ($r$ being the usual \sch coordinate) and $x = r + \log(r-1)$ is the tortoise coordinate. We will consider trajectories for which the particle falls into the BH,  $x_\mathrm{P} \rightarrow - \infty$ and $r_\mathrm{P} \rightarrow 1$.  Using Eq.~\eqref{eq:veloc} we find that the trajectory of the particle obeys $x_\mathrm{P} \simeq -t$, $\phi_\mathrm{P} \simeq 0$ (where we have chosen the constants of integration to vanish). 
In this article we will also need the next-to-leading order corrections to this trajectory as $t \rightarrow \infty$. This is easily found from Eq.~\eqref{eq:veloc}:
\begin{align} \label{eq:trajPPPlunge}
    x_\mathrm{P} &\simeq -t - \frac{L^2+1}{2E^2} e^{-t -1} + \mathcal{O}(e^{-2t}) \; , \\
    \phi_\mathrm{P} &\simeq - \frac{L}{E} e^{-t-1} + \mathcal{O}(e^{-2t}) \; .
\end{align}

\subsection{Waveform in time-domain from the Green function} \label{sec:GreenTimeDomain}
After decomposition in spherical harmonics, linear gravitational perturbations $\psi_{\ell m}(x,t)$ of the \sch geometry obey the Regge-Wheeler-Zerilli (RWZ) equations~\cite{Regge:1957td, Zerilli:1970aa},
\begin{equation}
   - \frac{\partial^2 \psi}{\partial t^2} + \frac{\partial^2 \psi}{\partial x^2} - V \psi = - S \label{eq:RWZeq} \; ,
\end{equation}
where we have suppressed the $\ell m$ indices for clarity, $V(x)$ is the Regge-Wheeler (odd) or Zerilli (even) potential, and $S$ is a source term modeling a point-particle on a geodesic $x=x_\mathrm{P}(t)$ given by (c.f.~Eqs.~23-24 in Ref.~\cite{Nagar:2006xv})
\begin{equation} \label{eq:Sourcefg}
    S( x,  t) = f( x_\mathrm{P}(t)) \delta( x -  x_\mathrm{P}( t)) + g( x_\mathrm{P}( t)) \delta'( x -  x_\mathrm{P}( t)) \; ,
\end{equation}
where the functions $f$ and $g$ depend on the energy and angular momentum of the point-particle and can be found e.g. in Refs.~\cite{Nagar:2006xv,DeAmicis:2025xuh,Kuntz:2025gdq}. For convenience, we write them in Appendix~\ref{app:source}.

For an observer at a distance $x$ greater than the source position $\bar x$, the Laplace-transformed Green function (defined to be the solution of Eq.~\eqref{eq:RWZeq} with a source $S = \delta(t-\bar t) \delta(x-\bar x)$) reads~\cite{Leaver:1986gd}
\begin{equation} \label{eq:GreenFarObserver}
    \tilde G(x, \bar x, \omega) =  i\frac{ \psi_H^-(\bar x, \omega) \psi_\infty^+( x, \omega)}{2 \omega A_\mathrm{in}} \; ,
\end{equation}
where $\psi_H^-$ ($\psi_\infty^+$) is the solution to the homogeneous RWZ equation~\eqref{eq:RWZeq} with ingoing (outgoing) boundary condition at the horizon (infinity), while $A_\mathrm{in}$ is the ingoing amplitude explicitly defined as
\begin{equation} \label{eq:defConnectionCoeffs}
    \psi_H^-(x, \omega) = A_\mathrm{in}(\omega) \psi_\infty^-(x, \omega) +  A_\mathrm{out}(\omega) \psi_\infty^+(x, \omega) \; ,
\end{equation}
where $\psi_\infty^-$ is the homogeneous solution ingoing at infinity. By definition, Eq.~\eqref{eq:GreenFarObserver} provides solutions of the non-homogeneous equation~\eqref{eq:RWZeq}, which are outgoing (ingoing) at infinity (horizon). The analytic properties in the complex $\omega$ plane of $\psi_H^-$ and $\psi_\infty^+$ were described in numerous references~\cite{Leaver:1986vnb,Arnaudo:2025uos,Kuntz:2025gdq, Arnaudo:2025kit,Motohashi:2026mbn}: $\psi_H^-$ has an infinite series of simple poles at negative imaginary frequencies $\omega^H_n = -i(n+1)/2$ ($n \geq 0$) that we will denote either \textit{horizon modes}, \textit{redshift modes} or \textit{Matsubara frequencies}, the latter name being adopted by analogy with finite-temperature field theory since these frequencies are multiples of the inverse Hawking temperature~\cite{Arnaudo:2025uos}. 
On the other hand, $\psi_\infty^+$ has a branch point and a pole at $\omega =0$.  Since $ A_\mathrm{in}(\omega)$ is proportional to the Wronskian between $\psi^-_H$ and $\psi^+_\infty$, it inherits the structure of both of these functions so it has poles at the horizon modes $\omega^H$,  a branch point and a pole at $\omega=0$, and additionally it has zeros at the QNM frequencies $\omega_{q}$, where $q=\{n\mathfrak{m} \}$, $n$ labels the overtone number and $\mathfrak{m}=\pm$ designates the sign of the real part of $\omega$ (QNM come in pairs with opposite signs of the real part of the frequency). Thus, the Green function defined in Eq.~\eqref{eq:GreenFarObserver} only has poles at the QNM frequencies, and a branch point at $\omega=0$~\cite{Leaver:1986gd}. 
Notice that a well-known exception occur for the Regge-Wheeler equation at the algebraically special frequency $\omega = -4i$ for the $n=8$ overtone, for which the zero in $A_\mathrm{in}$ at the QNM value is cancelled by a pole in $A_\mathrm{in}$ at the corresponding Matsubara mode~\cite{MaassenvandenBrink:2000iwh,Kubota:2026hdv}, so that $A_\mathrm{in}$ is regular (no zero, no pole) at $\omega=-4i$.
The inverse Laplace transform of Schwarzschild's Green function was computed in recent works~\cite{Arnaudo:2025uos,Arnaudo:2025kit,Su:2026fvj}. Note that we focus here on a specific $\ell$ harmonic, but the full \sch Green function can be recovered only after summing on all these modes~\cite{Casals:2009zh,Casals:2013mpa,Casals:2019heg,Aruquipa:2026tga,Aruquipa:2026kqk}.
Since redshift modes are excited when the particle approaches the horizon, we assume that $\bar x$ is situated to the left of the ``bounce radius"~\cite{Arnaudo:2026tcy}, i.e. $\bar x<0$.
In this case the time-domain Green function can be obtained \textit{at all times} by closing the integration contour defining it either on the upper or on the lower half of the complex plane, the switch between the two contours being given by the causal time $t - x = \bar t - \bar x$. Observe that, since we restrict to a single $\ell$ harmonic, our ``causal time" does not necessarily represent propagation on a \sch geodesic~\cite{Aruquipa:2026tga}; we will nonetheless use the terminology ``causality" or ``light-cone" to designate the characteristic lines $t - x = \bar t - \bar x$. 
Hence, we write the time-domain Green function as
\begin{align} \label{eq:GreenTimeDomainGeneric}
    G(x, \bar x, t-\bar t) &= i  \Theta(t-\bar t - x + \bar x) \int_{i \gamma - \infty}^{i \gamma+\infty} \frac{\mathrm{d}\omega}{2\pi} \nonumber \\
    &\times  \frac{e^{-i \omega(t-\bar t)}}{2 \omega A_\mathrm{in}(\omega)} \psi_H^-(\bar x, \omega) \psi_\infty^+( x, \omega) \; ,
\end{align}
where the Heaviside function $\Theta$ imposes the vanishing of the Green function outside of the light-cone, and $\gamma>0$. Here, we voluntarily left explicit the integral over frequencies $\omega$, bearing in mind that the contour can be closed on the lower half-plane as soon as $t - x \geq \bar t - \bar x$. In the following, it will sometimes be useful to approximate the Green function using the residue theorem as a QNM sum, neglecting the pole and branch cut at $\omega=0$. This approximation can be written as
\begin{align}
    &G(x, \bar x, t-\bar t) \simeq   \Theta(t-\bar t - x + \bar x) \nonumber \\
    &\times\sum_{q}  \frac{e^{-i \omega_{q}(t-\bar t)}}{2 \omega_{q} \alpha_{q}} \psi_H^-(\bar x, \omega_{q}) \psi_\infty^+( x, \omega_{q}) \; , \label{eq:GreenApproxQNM}
\end{align}
where the sum runs over overtones and mirror modes  $q=\{n\mathfrak{m} \}$, and the $\alpha_{q}$ are given by the expansion of the ingoing amplitude near to the QNM frequencies, $A_\mathrm{in}(\omega) \simeq \alpha_{q}(\omega-\omega_{q})$. However, we will make our results as general as possible and avoid to neglect the branch cut when possible.

If, on the other hand, we neglect the initial conditions, we can find the field generated by a plunging point-particle just by convolving the Green function with the source~\eqref{eq:Sourcefg}, which gives:
\begin{align}\label{eq:psi_inf}
    \psi(x,t) &= i \int \frac{\mathrm{d}\omega}{2\pi} \frac{e^{-i \omega u}}{2 \omega A_\mathrm{in}} \int_{-\infty}^{\infty} \mathrm{d} \bar t \int_{-\infty}^{\infty} \mathrm{d} \bar x \,  \Theta(t-\bar t - x + \bar x) \nonumber \\ 
    &\times e^{i \omega \bar t} \psi_H^-(\bar x, \omega) S(\bar x, \bar t) \; ,
\end{align}
where we have approximated $ \psi_\infty^+( x, \omega) \simeq e^{i \omega x}$ for an observer at large distances, and $u=t-x$. 
We use the delta-functions to cancel the integral in $\bar x$. Since there is also a derivative of a delta-function in the source, there appears two components in the waveform~\cite{DeAmicis:2025xuh,Kuntz:2025gdq}: an ``impulsive" (or ``direct", or ``instantaneous") contribution, coming from the derivative of the $\Theta$ function and involving only the source on the light-cone $\bar t - \bar x = u$; and an ``activation" (or ``historical") piece, involving an integral of the source at retarded times. They are given by 
\begin{align}
    &\psi = \psi_I + \psi_A \; , \nonumber \\
    &\psi_I(x,t) = - \frac{i g(x_\mathrm{P}(\tilde t))}{1- v_\mathrm{P}(\tilde t)} \int \frac{\mathrm{d}\omega}{2\pi} \frac{e^{-i \omega u}}{2 \omega A_\mathrm{in} } e^{i \omega \tilde t} \psi_H^-( x_\mathrm{P}(\tilde t), \omega) \; , \nonumber  \\
    &\psi_A(x,t) = i \int \frac{\mathrm{d}\omega}{2\pi} \frac{e^{-i \omega u}}{2 \omega A_\mathrm{in}} \int_{-\infty}^{\tilde t} \mathrm{d}\bar t \; e^{i \omega \bar t} \nonumber \\
    &\times \bigg( f(x_\mathrm{P}(\bar t)) \psi_H^-(x_\mathrm{P}(\bar t), \omega) - g(x_\mathrm{P}(\bar t)) \frac{\partial}{\partial x}\psi_H^-(x_\mathrm{P}(\bar t), \omega)  \bigg) \; , \label{eq:impulsiveActivation}
\end{align}
where we recall that $v_\mathrm{P} = \mathrm{d} x_\mathrm{P}/\mathrm{d}t$ and the time $\tilde t$ is defined as the retarded time at which a signal propagating on the light-cone (a direct wave) starting from the point-particle at $\tilde t$ reaches an observer at $t$:
\begin{equation} \label{eq:defTildet}
    \tilde t - x_\mathrm{P}(\tilde t) = u \ .
\end{equation}
Eq.~\eqref{eq:impulsiveActivation} will provide the basis for our investigation on redshift modes. We now examine its asymptotics when the particle falls into the BH.

\section{Leading order redshift mode} \label{sec:LO}

Since we are interested to the gravitational radiation emitted when the particle approaches the horizon, we isolate the contribution of the integrals in Eq.~\eqref{eq:impulsiveActivation} emerging in the limit $x_\mathrm{P}\to -\infty$. In this limit, $x_\mathrm{P}(\tilde t) \sim -\tilde t$, and $u \sim 2\tilde t\to +\infty$ (see Eq.~\eqref{eq:defTildet}). 
Furthermore, one has $f(x_\mathrm{P}) \simeq f_0 e^{x_\mathrm{P}}$, $g(x_\mathrm{P}) \simeq g_0 e^{x_\mathrm{P}}$ with $f_0, g_0$ two constants which depends on the energy $E$ and angular momentum $L$ of the particle (and $\ell$ and $m$ as well), but not on the frequency $\omega$. This behavior can be obtained from the expressions in Appendix~\ref{app:source} by noticing that $f$ and $g$ are proportional to the redshift factor $1-1/r$. 
Finally, $ \psi_H^-( x_\mathrm{P}, \omega) \simeq e^{-i \omega x_\mathrm{P}}$. 
Let us begin by analyzing the behavior of the impulsive term, which is the simplest to compute and interpret.

\subsection{Impulsive term} \label{sec:impulsive}

Recalling the definition of $\tilde t$ in Eq.~\eqref{eq:defTildet}, we get that in this limit the impulsive term is very simple:
\begin{equation}
    \psi_I(x,t) \simeq - \frac{i g_0}{1-v_P(\tilde t)} e^{x_P(\tilde t)} \int \frac{\mathrm{d}\omega}{2\pi} \frac{1}{2 \omega A_\mathrm{in}}  \; .
\end{equation}
Thus, the impulsive term consists in a pure redshift mode, which decays as a direct wave following the point-particle trajectory as $e^{x_\mathrm{P}}$. Notice that the integral involved in this formula is finite, as it corresponds to the value of the Green function~\eqref{eq:GreenTimeDomainGeneric} on the light-cone in this asymptotic regime.

To find the behavior of the impulsive term in the time-domain, we just use the relation $x_\mathrm{P}(\tilde t) \simeq - \tilde t \simeq -u/2$ in the asymptotic regime so that
\begin{equation} \label{eq:impulsiveLO}
    \psi_I(x,t) \simeq - g_0 e^{-u/2} \int \frac{\mathrm{d}\omega}{2\pi} \frac{1}{4 \omega A_\mathrm{in}}  \; .
\end{equation}
We see here the redshift mode appearing, with a decay given by the surface gravity of the BH $\kappa = 1/2$ (in units where $M=1/2$). 
There will be higher-order corrections to this formula which decay as higher multiples of the surface gravity, which we derive in Section~\ref{sec:proofAllOrders}.

\subsection{Activation term} \label{sec:activationLowestOrder}

To find the behavior of the activation term when the particle falls into the BH, we have to compute the integral in this limit. Let us choose a time $\bar t_1$ in the past for which the particle is already close enough to the BH, $-x_\mathrm{P}(t_1) \gg 1$, and neglect the contribution to the waveform coming from times $\bar t < \bar t_1$ (this contribution will contain QNMs, in which we are not interested here). Using the same approximations than in Section~\ref{sec:impulsive}, we get
\begin{align}
    \psi_A(x,t) &= i \int \frac{\mathrm{d}\omega}{2\pi} \frac{e^{-i \omega u} (f_0 + i \omega g_0)}{2 \omega A_\mathrm{in}} \nonumber \\
    &\times \int_{t_1}^{\tilde t} \mathrm{d}\bar t \; e^{i \omega ( \bar t - x_\mathrm{P}(\bar t)) + x_\mathrm{P}(\bar t)} \; .
\end{align}
Note that Ref.~\cite{Oshita:2025qmn} proposed to deal with this integral using a stationary phase approximation to find the redshift modes, called direct waves there. We disagree with this computation, because the stationary point (as computed in Ref.~\cite{Oshita:2025qmn}) is situated exactly at $\bar t = \tilde t$ and thus lies on the border of the domain of integration, hence one cannot use a stationary phase approximation. However, using a controlled expansion of the trajectory $x_\mathrm{P}(\bar t)$ as $\bar t \rightarrow \infty$ allows to straightforwardly perform the integral in $\bar t$. Indeed, using that at leading order $x_\mathrm{P}(\bar t) \simeq - \bar t$, we have to compute
\begin{equation}
    \int_{t_1}^{\tilde t} \mathrm{d}\bar t \; e^{\bar t (2i \omega -1) } = \frac{e^{\tilde t (2i \omega -1) } - e^{ t_1 (2i \omega -1) }}{2 i \omega - 1} \ .
\end{equation}
Using again that $\tilde t = u/2$ we find
\begin{align}
     \psi_A(x,t) &= i \int \frac{\mathrm{d}\omega}{2\pi} \frac{f_0 + i \omega g_0}{2 \omega (2 i \omega -1) A_\mathrm{in} } \nonumber \\
     &\times \big( e^{-u/2} - e^{-i \omega u} e^{ t_1 (2i \omega -1) } \big)  \; . \label{eq:psiALowestOrder}
\end{align}
The activation piece of the waveform is thus composed of a redshift mode $\propto e^{-u/2}$, and a piece oscillating as $e^{i \omega u}$. Looking at the integrand in Eq.~\eqref{eq:psiALowestOrder}, however, an interesting property manifests itself: it seems that a new pole at the Matsubara frequency $\omega_0^H = -i/2$ is present at the denominator. This new pole was called \textit{horizon mode} in Ref.~\cite{Zimmerman:2011dx}, and can \textit{a priori} give a new contribution to the waveform oscillating at the Matsubara frequency as advocated in Ref.~\cite{Zimmerman:2011dx}. Even though this is the same frequency than a redshift mode, this new spectral component is in theory distinct from a redshift mode because it involves the Green function evaluated at the Matsubara frequency and not at the QNM frequency like in Eq.~\eqref{eq:GreenApproxQNM}~\cite{Rosato:2026moe}. 

However, in our case the contribution of this new pole to the integral (closed in the lower half-plane) vanishes for at least two reasons. The first one is that the ingoing amplitude $A_\mathrm{in}$ itself has a pole at the Matsubara frequency, as mentioned in Section~\ref{sec:GreenTimeDomain}. Hence, $(2 i \omega -1) A_\mathrm{in}$ is regular at this frequency. 

Even letting aside this fact (for example, because one could be interested in beyond-GR theories where $A_\mathrm{in}$ could present a different analytic structure), it is easy to see that evaluating the residue at this new ``pole" gives a contribution $\propto e^{-u/2}-e^{-u/2}$ (cf. Eq.~\eqref{eq:psiALowestOrder}), and thus it vanishes. This was to be expected since, as mentioned in Section~\ref{sec:GreenTimeDomain}, we could have used from the very beginning a ``QNM sum" approximation to the Green function (Eq.~\eqref{eq:GreenApproxQNM}) where this fake pole would never have given a contribution to the residue theorem. 
Now, in order to gain physical insight, let us indeed use the residue theorem and pick up the QNM frequencies only, again ignoring the branch cut and pole at $\omega=0$. We get
\begin{align}
     \psi_A(x,t) &\simeq \sum_{q} \frac{f_0 + i \omega_{q} g_0}{2 \omega_{q} (2 i \omega_{q} -1) \alpha_{q} } \nonumber \\
     &\times \big( e^{-u/2} - e^{-i \omega_{q} u} e^{ t_1 (2i \omega_{q} -1) } \big)  \; . \label{eq:psiALowestOrderQNMApprox}
\end{align}
Eq.~\eqref{eq:psiALowestOrderQNMApprox} thus shows that $\psi_A$ is composed of a redshift mode $\propto e^{-u/2}$, and a QNM sum oscillating with frequencies $e^{-i \omega_{q} u}$. Notice that the redshift mode is given by the upper bound of the integral, again confirming its interpretation as a direct wave from the point-particle. On the other hand, the QNMs arise from the lower bound, and originate from a portion of the signal which is inside the light-cone.

\subsection{Total field}\label{sec:total_field}

Let us drop the lower bound of the integral in Eq.~\eqref{eq:psiALowestOrder}, which corresponds to the QNM component of the waveform. 
We find the total contribution to the field by summing the impulsive and activation pieces:
\begin{equation} \label{eq:totFieldLO}
    \psi(x,t) \simeq i e^{-u/2} \bigg( f_0 + \frac{g_0}{2} \bigg)   \int \frac{\mathrm{d}\omega}{2\pi} \frac{1}{2 \omega A_\mathrm{in} (2 i \omega-1)}  \; .
\end{equation}
Notice an apparently nontrivial simplification: the $g_0$ term no longer involves frequencies at the numerator as in Eq.~\eqref{eq:psiALowestOrder}. Instead, both $g_0$ and $f_0$ terms multiply a common integral. As mentioned in Section~\ref{sec:activationLowestOrder}, although the integrand seems to display a new pole at the Matsubara frequency $\omega_0^H = -i/2$, in fact it does not due to another pole in $A_\mathrm{in}$ at that frequency. 

Now comes a surprising property. The integrand in Eq.~\eqref{eq:totFieldLO} vanishes as $1/|\omega|^2$ as $|\omega| \rightarrow \infty$, because one has $A_\mathrm{in}(\omega) \rightarrow 1$ in this limit~\cite{PhysRevD.55.468}. On the other hand, let us recall that Jordan's lemma states that the large-arc integral vanishes as long as the integrand vanishes more quickly than $1/|\omega|$. Thus, we can conveniently close the contour of integration either in the lower or upper half-plane, and the value of the integral should be the same. Noting that the inverse Laplace transform is defined in Eq.~\eqref{eq:GreenTimeDomainGeneric} with $\gamma>0$ and that there are no poles nor branch cuts above the integration axis, we get that the amplitude of the lowest-order redshift mode in Eq.~\eqref{eq:totFieldLO} vanishes!

\begin{figure}
    \centering
    \includegraphics[width=\linewidth]{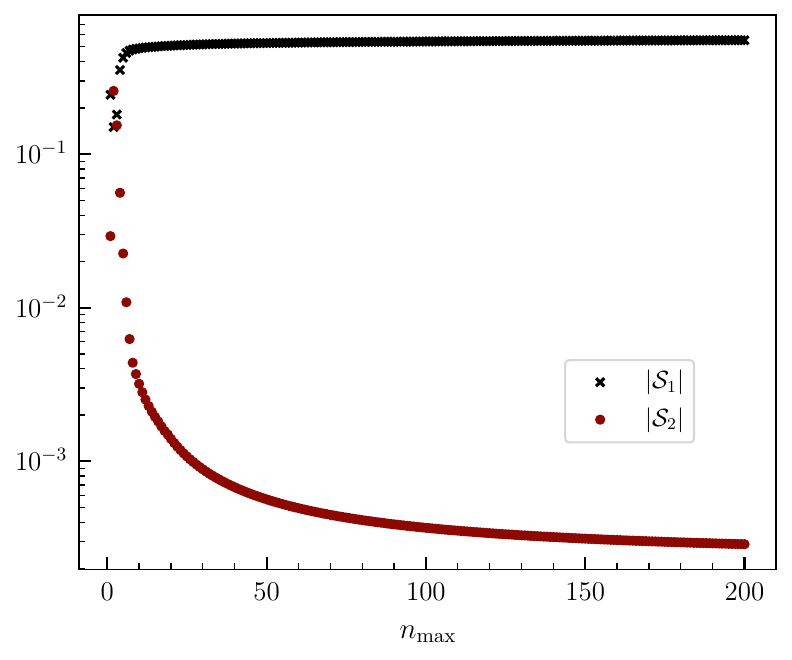}
    \caption{Approximated value of $|\mathcal{S}_1|$ and $|\mathcal{S}_2|$ as a function of the highest overtone $n_{\rm max}$ accounted for in the sum. Data are publicly available at Ref.~\cite{QNMData2026}.}
    \label{fig:s12}
\end{figure}

This result is surprising because intermediate steps of the computation contained quantities proportional to redshift modes that were non-vanishing. For example, the integrand of the impulsive term contribution to the redshift mode in Eq.~\eqref{eq:impulsiveLO} vanishes as $1/|\omega|$ as $|\omega| \rightarrow \infty$, and so Jordan's lemma is no longer applicable. In order to get an idea on the scaling of these integrals, let us use our QNM sum approximation and compute the amplitude of the impulsive term in Eq.~\eqref{eq:impulsiveLO}, $\mathcal{S}_1$, and the amplitude of the total field in Eq.~\eqref{eq:totFieldLO}, $\mathcal{S}_2$. Summing over the first $n=200$ overtones for the Regge-Wheeler equation, we get the approximated values
\begin{align}
    \mathcal{S}_1 &= \sum_{q} \frac{1}{2 \omega_{q} \alpha_{q}} \simeq 0.552 \; , \nonumber \\
    \mathcal{S}_2 &= \sum_{q} \frac{1}{2 \omega_{q} \alpha_{q} (2 i \omega_{q}-1)} \simeq -3 \times 10^{-4} \; ,\label{eq:redshiftAmplLOQNMApprox}
\end{align}
which are real due to the sum over mirror modes $\mathfrak{m} = \pm$. 
To obtain the values of $\alpha_q$, we have used the Mano-Suzuki-Takasugi method as explained in App.~\ref{app:MST}.  Note that the algebraically special frequency $\omega=-4i$ (corresponding to $n=8$) does not contribute to the sums as $A_{\rm in}$ does not have a zero there, as explained in Section~\ref{sec:GreenTimeDomain}.
Fig.~\ref{fig:s12} shows the convergence of $|\mathcal{S}_1|$ and $|\mathcal{S}_2|$ as more and more overtones are taken into account in the sum. 
Thus, we get the remarkable result that the impulsive term indeed decays as a redshift mode $e^{-u/2}$ with a nonzero amplitude as $u \rightarrow \infty$, and is \textit{exactly} the compensation needed to cancel the same redshift term present in the activation piece of the waveform. 

Before closing this section, let us note that it could seem inappropriate to evaluate the amplitude of redshift modes neglecting the branch cut in Eq.~\eqref{eq:redshiftAmplLOQNMApprox}, because we are in the asymptotic regime $u \rightarrow \infty$ where usually the tail contribution (originating from the branch cut) dominates the waveform. However, it is now clear that the redshift mode amplitude involves the Green function computed \textit{on the light-cone} $\bar t = \tilde t$. The results of Ref~\cite{Arnaudo:2025kit} suggest that the branch cut contribution to the Green function on the light-cone is small. Indeed, we recover this fact in our approach: the amplitude $\mathcal{S}_2$ is small but nonzero, suggesting that the branch cut contribution to the integral in Eq.~\eqref{eq:totFieldLO} (that should be added to $\mathcal{S}_2$ in order to make the integral exactly vanishing) is itself small. 
\section{Vanishing of all redshift modes} \label{sec:proofAllOrders}

Since the lowest order redshift mode, proportional to $e^{-u/2}$, is vanishing, we are naturally led to consider the question of higher-order modes: do they vanish as well? Indeed, it was argued in Ref.~\cite{Zimmerman:2011dx} that, although the leading order horizon mode is zero, the next-to-leading order (proportional to $e^{-u}$ for a \sch BH) is nontrivial. However, in this Section we shall prove that \textit{all} redshift modes, horizon modes, and direct waves vanish for a \sch BH. 
In essence, our proof relies on the following observations:
\begin{itemize}
    \item Redshift modes involve the Green function computed on the light-cone $\bar t - \bar x = t - x$ for $t -x \rightarrow \infty$;
    \item They also involve an integral of the source in time, which brings a factor $\omega^{-1}$;
    \item Although the Green function itself is discontinuous on the light-cone, the discontinuity being given by the integral $\int \mathrm{d} \omega (\omega A_\mathrm{in})^{-1}$ (see Eq.~\eqref{eq:GreenTimeDomainGeneric}), the redshift mode amplitude is continuous on the light-cone because of the supplementary $\omega^{-1}$ factor which allows to close the contour either on the upper or on the lower half-plane;
    \item By causality, this amplitude vanish.
\end{itemize}

However, the details of the proof are quite technical. Below, we first give a proof that redshift modes generated by the $f$ source term are vanishing, before going to the more delicate case of $g$.

\subsection{Term proportional to$f$ }

Let us first rewrite the term proportional to $f$ in Eq.~\eqref{eq:impulsiveActivation} using the change of variable $y = \bar t - x_\mathrm{P}(\bar t)$ in the integral:
\begin{align}
     \psi_A(x,t)\Big|_{f} &= i \int \frac{\mathrm{d}\omega}{2\pi} \frac{e^{-i \omega u} }{2 \omega A_\mathrm{in}} \int_{-\infty}^{u} \frac{\mathrm{d}y }{1-v_\mathrm{P}(y)} e^{i \omega \bar t(y)} \nonumber\\
     &\times f(x_\mathrm{P}(y)) \psi_H^-(x_\mathrm{P}(y), \omega) \; .
\end{align}
We now write $\psi_H^-(x, \omega) = e^{-i \omega x} \lambda(x, \omega)$. Close to the horizon $x \rightarrow - \infty$, the function $\lambda$ can be expanded as a power-series as
\begin{equation}
    \lambda(x, \omega) = \sum_{k \geq 0} \lambda_k(\omega) e^{k x} \; ,
\end{equation}
where $\lambda_0=1$ and the full series can be constructed recursively. For example, we get that $\lambda_1$ is given by
\begin{align} \label{eq:lambda1}
    \lambda_1 &=\frac{ 3 - \ell(\ell+1)}{e(2 i \omega-1)} \quad \mathrm{(Regge-Wheeler)} \; , \nonumber \\
    \lambda_1 &=- \frac{3 - 2 \ell - \ell^2 + 2 \ell^3 + \ell^4}{4 e (1 + \ell + \ell^2)(2 i \omega-1)} \quad \mathrm{(Zerilli)} \; .
\end{align} 
We see in Eq.~\eqref{eq:lambda1} the appearance of a pole at the lowest-order horizon mode frequency $\omega^H_0 = -i/2$, which we mentioned in Section~\ref{sec:GreenTimeDomain}. This supplementary pole makes the Green function regular at the horizon mode frequency. 
The most important property of the $\lambda_k$'s that we will use is that $\lambda_k = \mathcal{O}(1)$ when $|\omega| \rightarrow \infty$~\footnote{More precisely, one has $\lambda_k = \mathcal O(\omega^{-1})$ for $k \geq 1$, but a $\mathcal{O}(1)$ scaling is sufficient for our purposes.}, which can be shown by recurrence. Then, we have
\begin{align} \label{eq:termfAllOrders}
     \psi_A(x,t)\Big|_{f} &= i \int \frac{\mathrm{d}\omega}{2\pi} \frac{e^{-i \omega u} }{2 \omega A_\mathrm{in}} \int_{-\infty}^{u} \frac{\mathrm{d}y }{1-v_\mathrm{P}(y)} e^{i \omega y}\nonumber \\
     &\times f(x_\mathrm{P}(y)) \lambda(x_\mathrm{P}(y), \omega) \; .
\end{align}
As before, let us take a lower bound $u_1$ to the $y$ integral such that $u_1 \gg 1$. We perform an expansion of the integrand for large $y$. This will involve an expansion of the particle trajectory $x_\mathrm{P}(y)$, that we can easily find using the expression
\begin{equation}
    \frac{\mathrm{d}x_\mathrm{P}}{\mathrm{d}y} = \frac{\mathrm{d}x_\mathrm{P}}{\mathrm{d}t} \frac{\mathrm{d}t}{\mathrm{d}y} = \frac{v_\mathrm{P}}{1-v_\mathrm{P}} \; ,
\end{equation}
and taking the expression of $v_\mathrm{P}$ in Eq.~\eqref{eq:veloc}. Similarly, we can perform an expansion of $f(x_\mathrm{P})$ to all orders for $y \rightarrow \infty$, and this will involve constants depending on $\ell$, $m$, $E$ and $L$, but not on the frequency $\omega$. Overall, we write
\begin{equation}
    \frac{ f(x_\mathrm{P}(y)) \lambda(x_\mathrm{P}(y), \omega)}{1-v_\mathrm{P}(y)} = \sum_{k \geq 1} a_k(\omega) e^{-k y/2} \; ,
\end{equation}
where the $a_k(\omega)$ have the same asymptotic properties than $\lambda_k(\omega)$ for large $\omega$. Hence,
\begin{equation}
    \psi_A(x,t)\Big|_{f} = i \sum_{k \geq 1} \int \frac{\mathrm{d}\omega}{2\pi} \frac{e^{-i \omega u} }{2 \omega A_\mathrm{in}} \frac{a_k(\omega)}{i \omega - k/2} \big[  \; e^{(i \omega-k/2) y} \big]_{u_1}^u \; .
\end{equation}
We can always close the $\mathrm{d}\omega$ integral on the lower half-plane (because it involves an integrand in the region $t-x \geq \bar t - \bar x$), but the exponential factors prevent us to close it in the upper half-plane for the moment. Let us evaluate the contributions from the residues and branch cut in the lower half-plane. 
Even if $A_\mathrm{in}$ did not have a pole at the Matsubara frequency $\omega = -i k / 2$ (which is the case for the algebraically special frequency $\omega = -4i$ as mentioned in Section~\ref{sec:GreenTimeDomain}), it is easy to see that the contribution from the new residue at $\omega = -i k / 2$ (the horizon mode) vanishes because the lower bound $u_1$ cancels with the upper bound $u$. Hence, there only remains the ``usual" QNM poles and branch cut contributions to the $\mathrm{d}\omega$ integral. They give a contribution to $\psi(x,t)$ sourced by the lower bound $u_1$ of the integral, which however does not give any redshift mode. On the other hand, the upper bound gives a redshift mode (or direct wave) with a vanishing amplitude since, as we did in Section~\ref{sec:total_field}, one can  close the integral in the upper half-plane due to the supplementary $\omega^{-1}$ factor, as $a_k = \mathcal{O}(1)$ for large $|\omega|$.

\subsection{Term proportional to $g$}

This term is a little more tricky to deal with. On the one hand, a straightforward replacement $\psi_H^-(x, \omega) = e^{-i \omega x} \lambda(x, \omega)$ in the impulsive piece gives
\begin{align}
     \psi_I(x,t) &= - \frac{i g(x_\mathrm{P}(\tilde t))}{1- v_\mathrm{P}(\tilde t)} \int \frac{\mathrm{d}\omega}{2\pi} \frac{1}{2 \omega A_\mathrm{in} } \lambda( x_\mathrm{P}(\tilde t), \omega) \; .
\end{align}
On the other hand, by noticing that 
\begin{equation}
    \frac{\partial \psi_H^-}{\partial x} = - i \omega e^{-i \omega x} \lambda(x, \omega) + e^{-i \omega x} \frac{\partial \lambda  }{\partial x} \; ,
\end{equation}
and that the second term of this derivative gives rise to a vanishing redshift mode amplitude in the activation piece for the same reasons than for the $f$ term, we get that the only nontrivial piece of the $g$ term is
\begin{align}
     \psi_A(x,t)\Big|_{g} &= i \int \frac{\mathrm{d}\omega}{2\pi} \frac{e^{-i \omega u} }{2 \omega A_\mathrm{in}} \int_{-\infty}^{u} \frac{\mathrm{d}y }{1-v_\mathrm{P}(y)} e^{i \omega y} \nonumber \\
     &\times i \omega  g(x_\mathrm{P}(y)) \lambda(x_\mathrm{P}(y), \omega) \; .
\end{align}
Now the $\omega^{-1}$ factor is no longer sufficient to have the required scaling for the use of Jordan's lemma. However, a simple integration by parts gives
\begin{align}
     &\psi_A(x,t)\Big|_{g} = i \int \frac{\mathrm{d}\omega}{2\pi} \frac{1}{2 \omega A_\mathrm{in}} \bigg[ \frac{g(x_\mathrm{P}(\tilde t))}{1- v_\mathrm{P}(\tilde t)}  \lambda( x_\mathrm{P}(\tilde t), \omega) \nonumber   \\
     &- e^{-i \omega u }    \int_{-\infty}^{u} \mathrm{d}y \,  e^{i \omega y} \frac{\mathrm{d}}{\mathrm{d}y} \bigg( \frac{g(x_\mathrm{P}(y)) \lambda(x_\mathrm{P}(y), \omega)}{1-v_\mathrm{P}(y)} \bigg)     \bigg] \; ,
\end{align}
where as before we have discarded the lower bound of the integral in the boundary term, because it gives rise to QNMs and tail but not to redshift modes.  The second line provides an integral in the same form as Eq.~\eqref{eq:termfAllOrders} and thus it vanishes, while the first line is exactly cancelled by the impulsive term. This concludes the proof of the vanishing of redshift modes to all orders.

\section{Relation with excitation coefficients} \label{sec:excitationCoeffs}

The calculation we have performed crucially relied on causality to get a finite value for the QNM amplitudes. In previous works~\cite{Leaver:1986gd,Hadar:2009ip,Berti_2006,Berti:2006wq,Zhang:2013ksa,Sun:1988tz,DellaRocca:2025zbe}, these amplitudes or excitation coefficients were computed by just neglecting the causal structure of the Green's function and regulating the divergences which inevitably appeared in the computation. We are thus led to the following question: what is the relation between the regularization procedure used in Refs.~\cite{Leaver:1986gd,Hadar:2009ip,Berti_2006,Berti:2006wq,Zhang:2013ksa,Sun:1988tz,DellaRocca:2025zbe} and the vanishing of redshift modes?
In Ref.~\cite{DeAmicis:2025xuh}, this question was already touched upon: it was shown that the redshift modes correspond to divergences in the QNM amplitudes as $u \rightarrow \infty$. In this Section, however, we will emphasize the link between the regularization procedure and the calculation performed in Refs.~\cite{Leaver:1986gd,Hadar:2009ip,Berti_2006,Berti:2006wq,Zhang:2013ksa,Sun:1988tz,DellaRocca:2025zbe} and redshift mode amplitudes.
Following up on the results of Ref.~\cite{DeAmicis:2025xuh}, we show that the regularization employed in previous calculations ~\cite{Leaver:1986gd,Hadar:2009ip,Berti_2006,Berti:2006wq,Zhang:2013ksa,Sun:1988tz,DellaRocca:2025zbe} is not an \emph{ad hoc} prescription introduced to remove an unphysical divergence. Rather, it is a direct consequence of the vanishing of the redshift mode amplitudes.
Let us rewrite Eq.~\eqref{eq:psi_inf} in the limit $u\to \infty$. This limit is formally equivalent to neglect the causality condition, i.e. $  \Theta(t-\bar t - x + \bar x) \rightarrow 1$, and we can perform the same computation as in Refs.~\cite{Leaver:1986gd,Hadar:2009ip,Berti_2006,Berti:2006wq,Zhang:2013ksa,Sun:1988tz,DellaRocca:2025zbe}. 
We furthermore use our QNM sum approximation and neglect the branch cut. Hence,
\begin{equation}\label{eq:psi_inf_QNM}
    \psi(x,t)\simeq \sum_{q}\frac{e^{-i\omega_q u}}{2\omega_q \alpha_q}\int {\rm d}\bar x \ \psi^{-}_H(\bar x,\omega_q)\hat S(\bar x,\omega_q) \ ,
\end{equation}
where we have defined
\begin{equation}
    \hat S(\bar x,\omega_q)=\int {\rm d}\bar t \ e^{i\omega_q \bar t}S(\bar x,\bar t) \ ,
\end{equation}
i.e. the Fourier transform of Eq.~\eqref{eq:Sourcefg}.
Eq.~\eqref{eq:psi_inf_QNM} is in the form
\begin{equation}\label{eq:psi_Cq}
\psi(x,t)\simeq \sum_{q} C_q \  e^{-i\omega_q u} \ ,
\end{equation}
where $C_q$ are constant complex numbers known as \emph{excitation coefficients}~\cite{Leaver:1986gd} and defined by
\begin{equation}\label{eq:Cq}
    C_q=\frac{A_{\rm out}}{2 \omega_q\alpha_q}\int_{\mathbb{R}} {\rm d} \bar x \ \frac{\psi^-_H(\bar x,\omega_q)\hat S(\bar x,\omega_q)}{A_{\rm out}} \ .
\end{equation}
As first noticed by Leaver~\cite{Leaver:1986gd}, the integral in Eq.~\eqref{eq:Cq} does not converge for point-particle sources apart for the fundamental mode.
Indeed, not only $\psi_H^-\to (r-1)^{-i\omega_q }$ for $r\to 1$, but also the source $\hat S$ diverges as $(r-1)^{1-i\omega_q }$ for $r\to 1$~\cite{Zhang:2013ksa}. Putting all together, the integrand of Eq.~\eqref{eq:Cq} (changing variable from $x$ to $r$) can be expanded in a Frobenius series as
\begin{equation}\label{eq:iq_exp}
\psi_H^-(\bar r, \omega_q) \hat{S}(\bar r, \omega_q)\bar r(\bar r-1)^{-1}\sim (\bar r-1)^{-2i\omega_q}\sum_{j}\xi_j (\bar r-1)^j \ ,
\end{equation}
where the coefficients $\xi_j$ come from the expansion of the source and wavefunction close to the horizon. 
This shows that the integrand has a non-integrable singularity when ${\rm Im}[\omega]<-1/2$.
To isolate this divergence, we split the integral in Eq.~\eqref{eq:Cq} as
\begin{align}\label{eq:Cq_exp}
    C_q=&\frac{1}{2 \omega_q\alpha_q}\int_{1}^{1+\epsilon} {\rm d}  \bar r\ \frac{\bar r}{\bar r-1} \psi^-_H(\bar r,\omega_q)\hat S(\bar r,\omega_q)  \nonumber\\
    +&\frac{1}{2 \omega_q\alpha_q}\int_{1+\epsilon}^{\infty} {\rm d} \bar r \ \frac{\bar r}{\bar r-1}  \ \psi^-_H(\bar r,\omega_q)\hat S(\bar r,\omega_q) \ ,
\end{align}
where $\epsilon$ is sufficiently small for the expansion in Eq.~\eqref{eq:iq_exp} to hold for $r\in [1,1+\epsilon)$. This is reminiscent of our choice of time for the lower bound of the integral in Sections~\ref{sec:LO}-\ref{sec:proofAllOrders}. 
The first line of Eq.~\eqref{eq:Cq_exp} can be thus integrated analytically:
\begin{equation}\label{eq:div}
   \sum_{j=0}^\infty \frac{\xi_j}{j+1-2i\omega_q} (\bar r-1)^{j+1-2i\omega_q}\Bigg|_{1}^{1+\epsilon} \ .
\end{equation}
Assuming that $\epsilon$ is small but finite, the divergent part is given by evaluating Eq.~\eqref{eq:div} at $r=1$. We can thus define the \emph{regularized excitation coefficients} as
\begin{equation}
    \tilde C_q\equiv C_q-\beta_q  \ ,
\end{equation}
where 
\begin{equation}
    \beta_q= - \lim_{\bar r\to 1}  \sum_{j=0}^\infty \frac{\xi_j}{j+1-2i\omega_q} (\bar r-1)^{j+1-2i\omega_q} \ .
\end{equation}
From Eq.~\eqref{eq:Cq_exp}, we get
\begin{align}\label{eq:tildeCq}
    \tilde C_q=&\frac{1}{2 \omega_q\alpha_q}\int_{1+\epsilon}^{\infty} {\rm d} \bar r \ \frac{\bar r}{\bar r-1}  \ \psi^-_H(\bar r,\omega_q)\hat S(\bar r,\omega_q) \nonumber\\
  &+ \sum_{j=0}^\infty \frac{\xi_j}{j+1-2i\omega_q} \epsilon^{j+1-2i\omega_q} \ .
\end{align}
The second term on the right-hand side of Eq.~\eqref{eq:tildeCq} can be rewritten as
\begin{equation}
   \int_{1+\epsilon}^\infty {\rm d} r \ \frac{{\rm d}}{{\rm d} r}\sum_{j=0}^\infty\left(-\frac{b_j e^{-(r-1)}}{j+1-2i\omega_q} (r-1)^{j+1-2i\omega_q}\right)\,,
    \label{eq:bounds_funcs}
\end{equation}
for some coefficients $b_j$.
Eq.~\eqref{eq:bounds_funcs} is the standard term added to the integral in Eq.~\eqref{eq:Cq} to regularize the (otherwise divergent) excitation coefficients~\cite{Leaver:1986gd,Zhang:2013ksa,DellaRocca:2025zbe,Sun:1988tz}. 
Replacing $C_q=\tilde C_q+\beta_q$ in Eq.~\eqref{eq:psi_Cq}, we get (similarly to Eq.~53 in Ref.~\cite{DeAmicis:2025xuh})
\begin{equation}\label{eq:psi_tildeCq_betaq}
    \psi(u)\simeq \sum_{q} \tilde C_q e^{-i\omega_q u}+\beta_q e^{-i\omega_q u} \ .
\end{equation}
The coefficient $\beta_q$ contains the divergent part of $C_q$, and is thrown away in previous calculations of the excitation coefficients~\cite{Leaver:1986gd,Hadar:2009ip,Berti_2006,Berti:2006wq,Zhang:2013ksa,Sun:1988tz,DellaRocca:2025zbe}. 
We can however use the results of Section~\ref{sec:LO} to naturally regularize the divergence, because they are compatible with this Section in the limit $u \rightarrow \infty$.
Schematically, $r-1\simeq e^{-t}\simeq e^{-u/2}$ in the limit $u\to \infty$ (see Section~\ref{sec:LO}), and therefore 
\begin{equation}
    \beta_q e^{-i\omega_q u}\simeq - \sum_{j=0}^\infty \frac{\xi_j}{j+1-2i\omega_q}e^{-(j+1)u/2} \ , 
\end{equation}
which is a redshift mode. Hence, our calculation provides a justification to the regularization procedure used in Refs.~\cite{Leaver:1986gd,Hadar:2009ip,Berti_2006,Berti:2006wq,Zhang:2013ksa,Sun:1988tz,DellaRocca:2025zbe}: since redshift modes vanish, we are allowed to just drop the divergent part of $C_q$ in order to get a finite value for asymptotic QNM amplitudes. 
We notice also that the contribution to redshift mode amplitude comes from a boundary term. It means that redshift mode must be direct wave sourced by the particle approaching the horizon as no information can come out from the horizon itself.

\section{Conclusions} \label{sec:CL}

We have shown that all redshift modes, exponentially decaying with frequencies given by integer multiples of the surface gravity of a BH, have vanishing amplitudes in waveforms generated by point particles plunging into \sch BHs. The same conclusion applies to the closely related direct waves and horizon modes. Another interesting outcome of our analysis is that the so-called impulsive contribution to the waveform, introduced in Refs.~\cite{DeAmicis:2025xuh,DeAmicis:2026wqd}, is in fact precisely the quantity required to cancel the redshift-mode contribution appearing in the activation piece of the waveform, as demonstrated in Section~\ref{sec:proofAllOrders}. The impulsive contribution can therefore be interpreted as a counterterm required for the consistency of the waveform.

On the one hand, the vanishing of all redshift modes implies that it will not be possible to probe near-horizon physics using these kind of signals. On the other hand, this property appears as a remarkable feature of BHs, somewhat analogous to the vanishing of their Love numbers~\cite{Rodriguez:2026iot}. A natural question is whether this property extends to more general situations, such as Kerr BHs or modified theories of gravity. 
Although extending our analysis to these cases is technically nontrivial and left for future work, it is worth emphasizing that the fundamental ingredient underlying the vanishing of redshift modes is causality. We therefore expect this feature to persist in a much broader class of scenarios. To make this discussion more concrete, let us imagine that, due to a modification of the effective potential, $A_\mathrm{in}$ no longer possesses poles at the Matsubara frequencies, or instead develops poles at frequencies different from $\omega_n^H$. Our results would nevertheless remain unchanged, since the proof presented in Section~\ref{sec:proofAllOrders} does not rely in any essential way on the existence or location of poles of $A_\mathrm{in}$.

Our calculation is consistent with the results of Refs.~\cite{DeAmicis:2025xuh,DeAmicis:2026wqd}, because redshift modes vanish only after summing all overtones in the waveform as shown in Section~\ref{sec:LO}. They still give a nonzero contribution to each overtone amplitude which is the reason for the divergence observed in Refs.~\cite{DeAmicis:2025xuh,DeAmicis:2026wqd}. On the other hand, we disagree with the applicability of the stationary phase approximation performed in Ref.~\cite{Oshita:2025qmn} because the stationary point lies on the border of the integration domain, as explained in Section~\ref{sec:LO}. The additional pole arising at the Matsubara frequencies when integrating the source term has previously been argued to generate the so-called ``horizon modes'' for Kerr BHs~\cite{Mino:2008at,Zimmerman:2011dx}. Our analysis shows, however, that no such horizon modes are present for \sch BHs. Our result hence seem to contradict the conclusions reached in Refs.~\cite{Oshita:2025qmn,Mino:2008at,Zimmerman:2011dx} applied to non-rotating BHs.  
Although we suspect that the redshift modes, horizon modes, and direct waves may also vanish in the Kerr case, determining whether rotating BHs can support nontrivial horizon modes remains an open question that we leave for future work.

Our results also provide a key building block toward the definition of time-dependent QNM amplitudes that asymptote to finite constants at late times for all overtones. 
Indeed, our calculation suggests a natural subtraction scheme in which redshift-mode contributions are removed directly at the level of the QNM amplitudes, thereby rendering them finite. The resulting regularization procedure is closely related to earlier regularization prescriptions for QNM amplitudes introduced in~\cite{Leaver:1986gd,Hadar:2009ip,Berti_2006,Berti:2006wq,Zhang:2013ksa,Sun:1988tz,DellaRocca:2025zbe}, as discussed in Section~\ref{sec:excitationCoeffs}. This construction should ultimately enable more consistent ringdown models, in which time-dependent amplitudes smoothly connect the inspiral and ringdown regimes through their gradual buildup toward finite asymptotic values at late times.

\section*{Acknowledgments}

We thank Enrico Cannizzaro, Gregorio Carullo, Marina De Amicis, Leonardo Gualtieri, Joachim Pomper, Romeo Felice Rosato and Laura Sberna for discussions.
A.K  thanks the Fundação para a Ciência e Tecnologia (FCT), Portugal, for the financial support to the Center for Astrophysics and Gravitation (CENTRA/IST/ULisboa) through grant No. UID/PRR/00099/2025 and grant No. UID/00099/2025, as well as to the FCT project ``Gravitational waves as a new probe of fundamental physics and astrophysics'' grant agreement 2023.07357.CEECIND/CP2830/CT0003.
M.D.R.~is supported by the MUR FIS2 Advanced Grant ET-NOW (CUP:~B53C25001080001) and by the INFN TEONGRAV initiative.

\appendix
\section{Mano-Suzuki-Takasugi method for Regge-Wheeler equation} \label{app:MST}
The Mano-Suzuki-Takasugi (MST) method provides series representation of the solutions of the Teukolsky-Sasaki-Nakamura equation \cite{Sasaki:2003xr,Mano:1996vt} and the Regge-Wheeler equation~\cite{Mano:1996mf,Casals:2015nja}. Remarkably, these series representations allow for a semi-analytical computation of 
\begin{equation}\label{eq:alphaq}
    \alpha_q=\frac{\mathrm{d} A_{\rm in}}{\mathrm{d}\omega}\Bigg|_{\omega=\omega_q} \ ,
\end{equation} 
which appears in Eq.~\eqref{eq:redshiftAmplLOQNMApprox}. 
Although we may directly use the MST method for Regge-Wheeler equation~\cite{Mano:1996mf,Casals:2015nja}, we adapted the code used in Ref.~\cite{DellaRocca:2025zbe}, which computes the amplitudes of Sasaki-Nakamura equation. Indeed, even if the latter can describe the response of a BH with generic spin, in the limit of non-rotating BHs, the Sasaki-Nakamura equation coincides with the Regge-Wheeler equation~\cite{Sasaki:1994aa}. 
The Sasaki-Nakamura amplitudes can then be computed from the Teukolsky amplitudes~\cite{Zhang:2013ksa,Berti_2006,DellaRocca:2025zbe}. 

\subsection{Teukolsky amplitudes}
\label{sec:RenAngMom}
The Teukolsky equation describes the response of a BH with mass $M$ and angular momentum $J=M a$ to a generic perturbation with spin $s$.
The MST method relies on the introduction of an additional parameter in the Teukolsky equation, the so-called \emph{renormalized angular momentum} $\nu$. This parameter enables the construction of a convergent series representation of the solution, obtained by expanding it in a basis of hypergeometric functions~\cite{Mano:1996vt}.
A solution of the homogeneous Teukolsky equation with purely ingoing boundary conditions at the horizon $r=r_+$ can be thus written as
\begin{equation}
    R^\nu_{\rm in}=e^{i\omega b \hat x}(-\hat x)^{-s-i(\omega+\tau)/2}
(1-\hat x)^{i(\omega-\tau)/2}p_{\rm in}^{\nu}(\hat x) \ ,\label{eq:R_hypergeometric}
\end{equation}
where 
\begin{equation}
    p^\nu_{\rm in}(\hat x)=\sum_{n=-\infty}^\infty a_{{\rm in},\,n}^\nu \ p_{n+\nu}(\hat x)\,,
    \label{eq:seriesp}
\end{equation}
and
\begin{align}
 p_{\nu+n}(\hat x)=F(n+\nu&+1-i\tau,-n-\nu-i\tau;\\
    &1-s-i \omega-i\tau;\hat x)\ ,\nonumber
\end{align}
with $F(a,b;c;z)$ the ordinary hypergeometric function~\cite{Abramowitz1972}. 
Hereafter we omit the dependence of the Teukolsky and Sasaki-Nakamura functions on $\ell,m$ and $\omega$, and we have defined $\hat x=(r_+-r)/b$, $r_+=(1+b)/2$, $\tau=(\omega -2ma)/b$, with $b=\sqrt{1-4 a^2}$. 
The coefficients $a_n$ in Eq.~\eqref{eq:seriesp} are determined by a three-term recurrence relation:
\begin{equation}
    \alpha_{n}^{\nu} a_{{\rm in},\,n+1}^{\nu}+\beta_n^{\nu} a_{{\rm in},\,n}^{\nu}+\gamma_n^{\nu} a_{{\rm in},\,n-1}^{\nu}=0 \ ,\label{eq:nu_3term}
\end{equation}
where
\begin{eqnarray}\label{alpha}
\alpha_n^\nu&=&\frac{i \omega b(n+\nu+1+s+i \omega)(n+\nu+1+s-i \omega)}{(n+\nu+1)(2n+2\nu+3)(n+\nu+1+i\tau)^{-1}}\,,\nonumber\\ 
\label{beta}
\beta_n^\nu&=&-\lambda-s(s+1)+(n+\nu)(n+\nu+1)+ \omega^2\,\nonumber\\
&+& \omega(\omega-2ma)
+\frac{ \omega(\omega-2ma)(s^2+ \omega^2)}{(n+\nu)(n+\nu+1)}\,,\nonumber\\
\label{gamma}
\gamma_n^\nu&=&-\frac{i \omega b(n+\nu-s+i \omega)(n+\nu-s-i \omega)}{(n+\nu)(2n+2\nu-1)(n+\nu-i\tau)^{-1}}\,.
\end{eqnarray}
This three term relation can in principle be used to determine the renormalized angular momentum $\nu$ for which the series \eqref{eq:seriesp} is convergent, however we empirically found that the ``monodromy method''~\cite{Nasipak:2024icb} provides more stable values at high overtones. 
To compute $\alpha_q$ in Eq.~\eqref{eq:alphaq}, we need the incident asymptotic amplitudes $A_{\rm inc}^T$ of the Teukolsky function $R_{\rm in}^\nu$. While the series representation in Eq.~\eqref{eq:seriesp} cannot directly be used to compute these amplitudes as the series does not converge at spatial infinity, we can find an analytic continuation of $R_{\rm in}^\nu$ at $r=\infty$ making use of an expansion of the Teukolsky function in terms of Coulomb wave functions~\cite{Mano:1996vt,Sasaki:2003xr}: 
\begin{equation}
R^\nu_{\rm in}=K_{\nu}R_{{\rm C}}^{\nu}+K_{-\nu-1}R_{{\rm C}}^{-\nu-1} \ ,
\label{eq:secondRin}
\end{equation}
where 
\begin{widetext}
\begin{eqnarray}
K_{\nu}
&=&\frac{e^{i\omega b}(2\omega b)^{s-\nu-N}2^{-s}i^{N}\Gamma(1-s-i \omega-i\tau)\Gamma(N+2\nu+2)}
{\Gamma(N+\nu+1-s+i \omega )\Gamma(N+\nu+1+i\tau)\Gamma(N+\nu+1+s+i \omega)}\nonumber \\
&&\times\left(\sum_{n=N}^{\infty}(-1)^{n}\frac{\Gamma(n+N+2\nu+1)}{(n-N)!}
\frac{\Gamma(n+\nu+1+s+i \omega)\Gamma(n+\nu+1+i\tau)}
{\Gamma(n+\nu+1-s-i \omega )\Gamma(n+\nu+1-i\tau)}a_{{\rm in},n}^{\nu}\right)\nonumber \\
&&\times\left(\sum_{n=-\infty}^{N}\frac{(-1)^{n}}{(N-n)!(N+2\nu+2)_{n}}
\frac{(\nu+1+s-i \omega )_{n}}{(\nu+1-s+i \omega )_{n}}a_{{\rm in},n}^{\nu}\right)^{-1} \ .
\end{eqnarray}
\end{widetext}
In Eq.~\eqref{eq:secondRin}, the function $R_{\rm C}^\nu$ is defined as
\begin{equation}
R^\nu_{{\rm C}}={\hat z}^{-1-s}\left(1-\frac{\omega b}{\hat z}\right)^{-s-i(\omega+\tau)/2}
f_{\nu}(\hat z) \ ,\label{eq:RC_definition}
\end{equation}
where 
\begin{align}
f_{\nu}(\hat z)= \sum_{n=-\infty}^{\infty}
(-i)^n\frac{(\nu+1+s-i \omega )_n}{(\nu+1-s+i \omega )_n}a^{\rm C}_n F_{n+\nu}(-is-\omega,\hat z) \ ,
\label{eq:series of Rc}\end{align} 
$\hat z=\omega (r- r_- )$, $r_-=(1-b)/2$, $(y)_{n}=\Gamma(y+n)/\Gamma(y)$, and $F_{N}(\eta,\hat z)$ is a Coulomb wave function, defined by 
\begin{align}
F_{N}(\eta,\hat z)=e^{-i\hat z}2^{N}&{\hat z}^{N+1}\frac{\Gamma(N+1-i\eta)}{\Gamma(2N+2)}\times\nonumber\\
&\Phi(N+1-i\eta,
2N+2;2i\hat z) \ ,
\label{eq:defcoulomb}
\end{align}
with $\Phi(\alpha,\beta;\hat z)$ the confluent hypergeometric function~\cite{Abramowitz1972}. 
The coefficients $a_{n}^{\rm C}$ satisfy the same three-term recurrence relation in Eq.~\eqref{eq:nu_3term} and thus the parameter $\nu$ is identical for both $R^\nu_{\rm in}$ and $R^\nu_{\rm C}$~\cite{Mano:1996vt}. 

Expanding Eq.~\eqref{eq:R_hypergeometric} at the horizon and Eq.~\eqref{eq:secondRin} at infinity, and matching with the asymptotic expansion 
\begin{equation}
    R^\nu_{\text{in}}\to \begin{cases}
B^{\text{trans}}\Delta^2 e^{-i(\omega-m a/r_+) r_\star} & r\to r_+\\
r^3 B^{\text{ref}} e^{i\omega r_\star} + r^{-1} B^{\text{inc}} e^{-i\omega r_\star} \quad & r\to +\infty 
\end{cases}\;,
\label{eq:RinAsy}
\end{equation}
where $r_\star$ is the tortoise coordinate in the Kerr background and $\Delta= r^2+a^2-r$,
we get 
\begin{subequations}
\begin{align}
B^{{\rm trans}}
=&\,b^{2s}e^{i  b (\omega+\tau)/2+\left(1+2\frac{\ln b}{1+b}\right)}
\sum_{n=-\infty}^{\infty}a_{{\rm in},n}^{\nu} \ ,\\
B^{{\rm inc}}
=&\,\omega^{-1}\left[K_{\nu}-ie^{-i\pi\nu}
\frac{\sin\pi(\nu-s+i \omega )}{\sin\pi(\nu+s-i \omega )}K_{-\nu-1}\right]\nonumber\\
&\,\times A_{+}^{\nu}e^{-i\left(\omega\ln\omega-\frac{1-b}{2}\omega\right)} \ , \\
B^{{\rm ref}}
=&\,\omega^{-1-2s}[K_{\nu}+ie^{i\pi\nu}K_{-\nu-1}]\nonumber\\
&\times A_{-}^
{\nu}e^{i\left(\omega\ln\omega-\frac{1-b}{2}\omega\right)} \ ,   
\end{align}
\label{eq:asymp_amp}
\end{subequations}
with

\begin{align}
A_{+}^{\nu}=&\,2^{-1+s-i \omega }e^{-\frac{\pi}{2}\omega}e^{\frac{\pi}{2}i(\nu+1-s)}\nonumber\\
&\frac{\Gamma(\nu+1-s+i \omega  )}{\Gamma(\nu+1+s-i \omega )}\sum_{n=-\infty}^{+\infty}a_{{\rm in},n}^{\nu} \ ,
\nonumber \\
A_{-}^{\nu}=&\,2^{-1-s+i \omega }e^{-\frac{\pi}{2}\omega}e^{\frac{-\pi}{2}i(\nu+1+s)}\nonumber \\
&\times \sum_{n=-\infty}^{+\infty}(-1)^{n}\frac{(\nu+1+s-i \omega )_{n}}{(\nu+1-s+i \omega )_{n}}
a_{{\rm in},n}^{\nu} \ .
\end{align}
\subsection{The Sasaki-Nakamura amplitudes}\label{sec:excitation_factors}
The normalized Sasaki-Nakamura amplitude $A_{\rm in}$ appearing in Eq.~\eqref{eq:alphaq}  can be computed from the Teukolsky amplitudes (with $s=-2$). 
In the same way than for the Teukolsky amplitudes, we can define $A^{\rm ref}$, $A^{\rm inc}$ and $A^{\rm trans}$ from the expansion 
\begin{equation}
    X^{\rm in}\to \begin{cases}
       A^{\rm trans} e^{-i(\omega-m a/r_+) r_\star}  &r \to r_+\\
        A^{\rm ref} e^{i\omega r_\star}+A^{\rm inc} e^{-i\omega r_\star} &r \to +\infty 
    \end{cases}\ .
    \label{eq:def_SN_in}
\end{equation}
The coefficients $A^{\rm ref}$, $A^{\rm inc}$ and $A^{\rm trans}$ are 
related to the amplitudes of the corresponding 
Teukolsky function $B^{\rm ref}$, $B^{\rm inc}$ and $B^{\rm trans}$ in Eqs.~\eqref{eq:asymp_amp} by (see e.g.~\cite{Sasaki:2003xr}):
\begin{subequations}\label{eq:amplitudesSN}
\begin{eqnarray}
    A^{\rm inc}&=& -4 \omega^2 B^{\rm inc}\ ,\\
    A^{\rm ref}&=& -\frac{c_0}{4 \omega^2} B^{\rm ref}\ \ ,\\
    A^{\rm trans}&=&d \ B^{\rm trans}\ ,
\end{eqnarray}
\end{subequations}
where 
\begin{align}
    d&=\sqrt{ 2M r_+} \left[8 M^2-12 i a m M-4 a^2 m^2\right.
    \nonumber\\
    &\left.+r_+\left(12 i a m+16 a m M
   \omega +24 i M^2 \omega -16 M\right)\right.\nonumber \\
   &\left.+r_+^2
   \left(-16 M^2 \omega ^2-24 i M \omega +8\right)\right]\,, 
\end{align}
and $c_0=-12 a \omega  (a \omega -m)+\lambda  (\lambda +2)-6i \omega$, with $\lambda=\ell(\ell+1)-s(s+1)$ for $a=0$ (see e.g.~\cite{DellaRocca:2025zbe} for the more involved case $a\ne 0$). 
The normalized amplitude $A^{\rm in}$ can be straightforward obtained via 
\begin{equation}
    A^{\rm in}=A^{\rm inc}/A^{\rm trans} \ .
\end{equation}
In the limit $a=0$, the Sasaki-Nakamura amplitudes coincide with the required Regge-Wheeler amplitudes.
The full dataset underlying Fig.~\ref{fig:s12} has been made publicly available at
Ref.~\cite{QNMData2026}.



\section{Source of the Regge-Wheeler-Zerilli equations} \label{app:source}
The functions $f$ and $g$ multiplying the delta-functions in the expression~\eqref{eq:Sourcefg} for the source term are given for the Zerilli equation by~\cite{Nagar:2006xv,Nagar:2005ea,DeAmicis:2025xuh,Kuntz:2025gdq}
\begin{align}
    g(r) &= - \frac{16 \pi Y_{\ell m}^* A(r)}{r E \mu (3 + r (\mu -2))} \big( L^2+r^2 \big) \label{eq:gfunc} \; , \\
    f(r) &=  -\frac{16 \pi Y_{\ell m}^* A(r)}{r E \mu (3 + r (\mu -2))} \bigg[ -2 i m L E v + \frac{5}{2} - \frac{r}{2}(\mu -2) \nonumber \\
    &+ \frac{L^2}{r^2} \big( 3 + r (m^2-\mu -2) \big) + \frac{L^2}{r^2(\mu -2)} \big( 3m^2-\mu -5 \big) \nonumber \\
    &+ \frac{r}{3+(\mu -2) r} \bigg( 6E^2 - A(r) (\mu -2) \bigg(1+\frac{L^2}{r^2} \bigg) \bigg)\bigg] \; ,
\end{align}
where $\mu = \ell(\ell+1)$ and we recall that $A(r) = 1 - 1/r$. We have normalized the point-particle mass to one. For the Regge-Wheeler equation, they are instead given by 
\begin{align}
    g(r) &= \frac{16 \pi \partial_\theta Y_{\ell m}^* A(r)}{r \mu (\mu - 2)}\frac{L}{E^2} \bigg( 1 + \frac{L^2}{r^2} \bigg) \ , \\
    f(r) &= - \frac{16 \pi \partial_\theta Y_{\ell m}^* A(r)}{r \mu (\mu - 2)} L \bigg( \frac{1}{2E^2r} \bigg( 1 - \frac{2L^2}{r} + \frac{3L^2}{r^2} \bigg) \nonumber \\
    &+ \frac{2L}{r} + i m \frac{v L^2}{r^2 E}  \bigg) \ .
\end{align}

\bibliography{refs2}

\end{document}